\ifdefined\pdfminorversion\pdfminorversion=4\fi
\ifdefined\pdfobjcompresslevel\pdfobjcompresslevel=0\fi
\documentclass[10pt,conference,a4paper]{IEEEtran}
\IEEEoverridecommandlockouts

\usepackage{cite}
\usepackage{amsmath,amssymb,amsfonts}
\usepackage{graphicx}
\usepackage{booktabs}
\usepackage{tabularx}
\usepackage{array}
\usepackage{microtype}
\usepackage{siunitx}
\usepackage{xcolor}
\usepackage{hyperref}
\usepackage{cleveref}
\usepackage{etc}

\definecolor{rqgreen}{HTML}{3F6B4F}
\providecommand{\rqbox}[2]{%
  \begin{tcolorbox}[
    colback=green!10,
    colframe=rqgreen,
    arc=1mm,
    boxrule=0.5pt,
    left=1mm,
    right=1mm,
    top=0.5mm,
    bottom=0.5mm,
    before skip=4pt,
    after skip=4pt
  ]
  \textbf{#1.} #2
  \end{tcolorbox}%
}
\definecolor{brandgreen}{HTML}{00915A}
\hypersetup{colorlinks=true,allcolors=brandgreen}
\newcommand{\best}[1]{\textbf{#1}}
\newcommand{\cmark}{\textcolor{brandgreen}{\textbf{Y}}}
\newcommand{\xmark}{\textcolor{black!35}{--}}

\newcolumntype{L}[1]{>{\raggedright\arraybackslash}p{#1}}
\newcommand{\tblsetup}{%
  \footnotesize%
  \setlength{\tabcolsep}{3pt}%
  \renewcommand{\arraystretch}{1.02}%
}

\newif\ifstudydone
\studydonetrue

\providecommand{\repourl}{https://github.com/FinAI-Lab/VERA}

\begin{document}

\title{VERA: A Responsible AI Dashboard for Assessing\\ LLM Deployments in Light of the EU AI Act\thanks{Accepted at the 33rd Asia-Pacific Software Engineering Conference (APSEC 2026), Software Engineering in Practice track, Bali, Indonesia, 7--10 December 2026. This is the authors' accepted version. \copyright~2026 IEEE. Personal use of this material is permitted. Permission from IEEE must be obtained for all other uses, in any current or future media, including reprinting/republishing this material for advertising or promotional purposes, creating new collective works, for resale or redistribution to servers or lists, or reuse of any copyrighted component of this work in other works.}}

\author{%
\IEEEauthorblockN{%
    Robin Qu\'eriaux\IEEEauthorrefmark{1},
    Solveig Bachellery\IEEEauthorrefmark{1},
    L\'ea D\'el\'eris\IEEEauthorrefmark{1},
    Saurabhsingh Rajput\IEEEauthorrefmark{2},\\
    Mariam Barry\IEEEauthorrefmark{1},
    Tushar Sharma\IEEEauthorrefmark{2},
    J\'er\^ome Lebecq\IEEEauthorrefmark{1}%
}
\IEEEauthorblockA{%
    \IEEEauthorrefmark{1}BNP Paribas, France\\
    \IEEEauthorrefmark{2}Dalhousie University, Halifax, Canada%
}%
}

\maketitle

\begin{abstract}
Institutions that deploy large language models in regulated settings often need technical evidence of responsible AI practices before approving a deployment.
COMPL-AI provides a technical interpretation of the EU AI Act that organises this evidence under six principle groups and eighteen technical requirements, twelve of which current benchmarks can measure.
The benchmark suites that produce this evidence are typically operated from a command line, while the staff who approve deployments are non-engineers.
We present VERA, an open-source self-hosted dashboard deployed to facilitate responsible AI processes within a large European bank. A reviewer selects a model and runs public benchmarks against it.
VERA scores the twelve requirements that current benchmarks can measure and instruments the energy consumed by each evaluation, adding run-level evidence for environmental impact.
We run four served open-weight models through the requirement pipeline, and the dashboard displays a score for each measurable requirement. A within-subject study then compares how participants read these benchmark results: from the raw output files in one condition, and through the VERA dashboard in the other.
The dashboard reduced the time needed to perform the responsible AI assessment, reducing the time required from 56 seconds using raw files to 30 seconds using our dashboard, on average.
Our dashboard scored highly (4.36 out of 5) on usefulness in the survey conducted within the organization.
The dashboard and the study material are released publicly to promote replicability and reproducibility in different organisational contexts.
\end{abstract}

\begin{IEEEkeywords}
Responsible AI, EU AI Act, Compliance tooling, User study, Software engineering in practice, Open source
\end{IEEEkeywords}

\section{Introduction}
\label{sec:intro}

Large language models are being built into products across almost every customer-facing function, from answering queries to triaging support requests and drafting documents.
Their outputs are not specified by deterministic rules, and generation can vary across runs when stochastic decoding is used.
They can also generate confident but false statements, known as hallucinations~\cite{ji2023hallucination,lin2022truthfulqa}, reproduce text from their training data~\cite{carlini2023extracting}, and answer the same question differently depending on the social group it names, such as a gender, an ethnicity or a religion~\cite{parrish2022bbq}.
Exact-output testing is insufficient for behaviour of this kind, since many prompts admit no single expected response against which every output can be checked.
Practitioners therefore evaluate such behaviour statistically, over benchmark suites that probe each failure mode~\cite{hendrycks2021mmlu,gehman2020realtoxicity}.

These risks have led regulators to impose documentation, testing and risk-management obligations rather than leave evaluation to good practice.
For high-risk AI systems, the EU AI Act requires technical documentation to be drawn up before the system is placed on the market or put into service and to demonstrate compliance with the applicable requirements~\cite{euaiact2024}.
Recital~27 separately recalls seven non-binding trustworthy-AI principles from the AI HLEG, namely human agency and oversight, technical robustness and safety, privacy and data governance, transparency, diversity, non-discrimination and fairness, societal and environmental well-being, and accountability~\cite{euaiact2024,hleg2019trustworthyai}.
For institutions subject to these obligations, producing and organising technical evidence becomes a routine engineering task before deployment proceeds.

\begin{figure*}[t]
  \centering
  \includegraphics[width=0.80\textwidth]{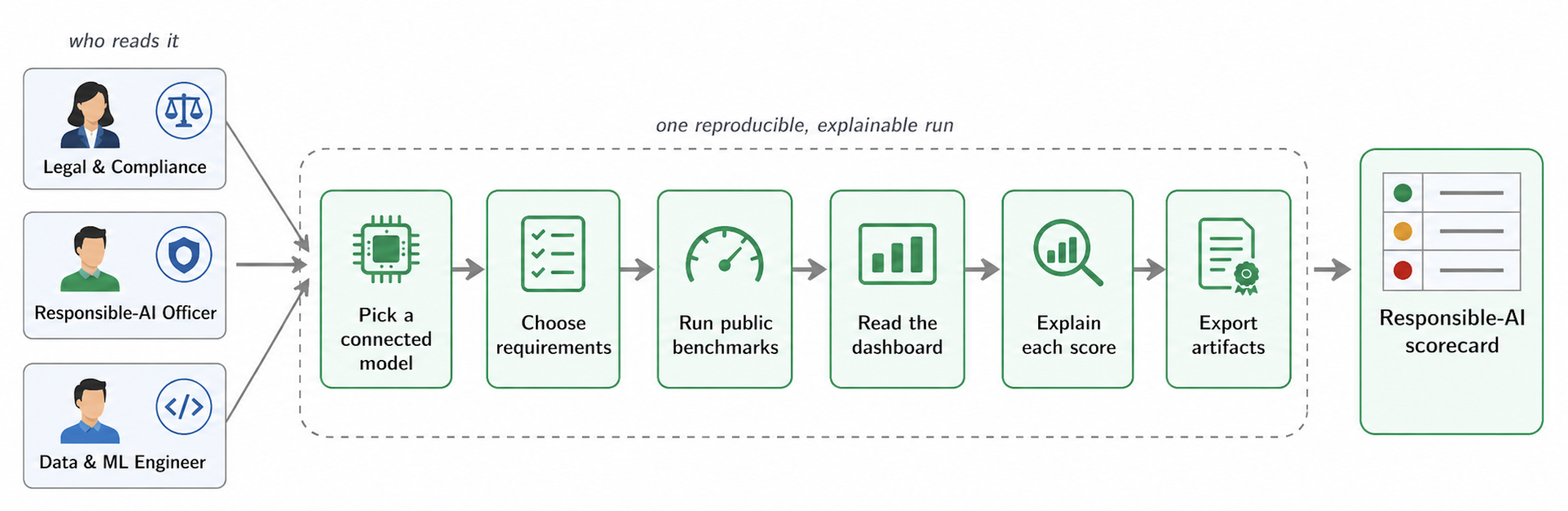}
  \caption{The role-agnostic path: select a model and requirement set, run public benchmarks, read a role-filtered outcome where each score opens, and export evidence tied to the run. 
  }
  \label{fig:journey}
\end{figure*}

In a regulated institution, however, the people who produce this evidence and the people accountable for the deployment are not the same.
In our institutional setting, approval rests with the model owner and business-oriented managers, while the evaluation suites are operated by engineers.
Supervision of the process is the responsibility of the second line of defence, in compliance or in risk management.
Over a year of preparing release decisions within responsible-AI processes at BNP Paribas, we observed three problems that recur because of this separation.
The first is access, since the suites used in these reviews are driven from a command line.
An engineer therefore runs the evaluation and brings a summary slide to the review meeting, where the reviewers cannot interrogate the numbers on it.
The second is opacity, since results arrive as a table of scores with no link to the tests behind them.
A reviewer who asks why a requirement scored poorly must match the run record against per-item outputs by hand.
In one review, a reviewer asked which benchmarks had produced an artificial-intelligence disclosure score of $0.50$ for one model, and answering meant scanning a 209-line harness log to isolate the ten self-disclosure responses behind that requirement, even though every fact needed was already in the files.
The third is traceability.
Release documentation is written after the run, so the filed document can become detached from the evaluation it describes as runs are repeated or updated.
All three are properties of the software that delivers the measurements rather than of the measurements themselves.

The measurement itself is already well supported.
COMPL-AI maps six principle groups in its technical interpretation to eighteen technical requirements and connects twelve of them to public benchmarks~\cite{guldimann2024complai}, while the suites beneath it fix which scenarios a report covers and how each benchmark is executed~\cite{liang2023helm,gao2021lmevalharness,derczynski2024garak,wang2023decodingtrust}.
Run against a served model, these tools write a run record, a harness log and the per-item outputs, which together hold the facts a reviewer needs.
The underlying suites, however, are designed primarily to produce and store measurements rather than to support the review workflow around them.
Working from their raw artifacts still requires technical tooling, requirement-level results are not consistently linked to the per-item evidence behind them, and downstream documentation need not record the exact run from which a value came.
Access, opacity and traceability therefore remain after measurement, and closing them is the first gap this work addresses.
A second gap lies inside the published taxonomy itself, since the remaining six requirements are handled without benchmark scores, and environmental impact is one of them.
It differs from the other five, because corrigibility and explainability have no agreed technical definition, whereas energy is a physical quantity for which the AI Act calls for lifecycle reporting standards and, for providers of general-purpose AI models, known or estimated model-energy documentation (Art.~40 Para~2, Annex~XI Para~2e)~\cite{euaiact2024}.
COMPL-AI collects that figure through a form about the resources used in training, although the energy an evaluation itself consumes can be instrumented directly on the host that runs it~\cite{rajput2024fecom}.

To close both gaps we built VERA (\textbf{V}erifiable \textbf{E}valuation for \textbf{R}esponsible \textbf{A}I), a self-hosted dashboard designed to let staff involved in deployment review produce, examine and file benchmark evidence in a browser without writing code.
It is deployed to facilitate responsible AI processes within the institution described above, and the path a reviewer follows is short.
A reviewer selects a served model, runs the public benchmark suites against it, and reads a score for each technical requirement.
Any score opens onto the benchmarks that produced it and the answers the model gave.
The export that leaves the review names the run behind it, so a later reader can check a figure rather than trust it.
VERA scores the twelve requirements the reference taxonomy can measure and adds instrumented run-level energy evidence to the environmental-impact requirement, while five of the eighteen remain on a documented human path.
\Cref{fig:journey} shows the route a reviewer takes, which needs no account, no configuration file and no code.
VERA has to produce those scores before their presentation can matter, so we evaluate both.
We first run four served open-weight models through five requirements at a benchmark sampling budget of 150, which shows that the tool recomputes its own aggregate end to end and separates the models it scores.
We then run a within-subject study administered by the dashboard itself, in which each of ten participants answers six factual questions about one completed evaluation from its raw files and six matched questions on VERA, while the application times each question and validates each answer.
The condition order is fixed and the participants are predominantly research staff rather than the compliance reviewers the tool targets, so we report a difference in reading performance rather than an isolated interface effect or a change in deployment decisions.

This paper makes the following contributions.

\begin{enumerate}
   \item We present VERA, an open-source self-hosted dashboard that turns a completed evaluation into requirement-level results a reviewer can read, expand to the tests and model responses behind them, and export with the identity of the run that produced them.
  \item We extend the evidence path to environmental impact by instrumenting the energy the served models consume during each evaluation run, replacing a written attestation for that run-level quantity with recorded evidence.
  \item We provide paired evidence that participants extract more correct facts from a requirement-oriented presentation than from the raw result files, using a within-subject study that the dashboard administers.
  \item We identify where that presentation fails, since aggregating benchmarks per requirement hides the coverage a reviewer must count, and a guided launch path does not by itself make a run reachable.
  \item We release the dashboard, the study instrument that produced these measurements, and the analysis script, available at \url{\repourl}.
\end{enumerate}

\clearpage
\section{Background and Related Work}
\label{sec:background}

Measuring a model and reading that measurement are separate problems, and the literature has focused primarily on the first.
This section reviews the work on each.

\subsection{Benchmark Suites and Requirement Taxonomies}
COMPL-AI is an independent effort by ETH Zurich, INSAIT and LatticeFlow AI that maps six principle groups in its technical interpretation to eighteen technical requirements and connects twelve of them to public benchmarks~\cite{guldimann2024complai}.
It is neither funded by the European Union nor part of the AI Act's harmonised standards, and its authors explicitly state that it is not official auditing software for legal compliance.
It publishes the technical interpretation, an open-source benchmarking suite and a public leaderboard, so both the taxonomy and the executable measurement layer are available to practitioners; its workflow remains evaluation-centred rather than organised around institutional review and filing.

The suites beneath that taxonomy execute the tests.
HELM fixes which scenarios a report covers~\cite{liang2023helm}, the language model evaluation harness fixes how a benchmark runs so that two runs are comparable~\cite{gao2021lmevalharness}, Garak probes for jailbreaks~\cite{derczynski2024garak}, and DecodingTrust bundles a trustworthiness suite~\cite{wang2023decodingtrust}.
Each answers a measurement question in isolation, and none reports its result against an Act-oriented technical requirement.
A separate line of research applies natural-language processing to the regulation itself, extracting obligations from legal text and checking artifacts against them~\cite{abualhaija2025llmregextraction,amaral2022nlpcompliance,abualhaija2024regchange}.
Deriving a requirement set is a different problem from operating one that already exists, and this paper addresses the second.

\subsection{Tool Support for Reviewing Results}
Running these suites against a served model yields a run record, a harness log and the per-item outputs, and \Cref{tab:related} compares what a reviewer can then do with that output.
The table shows a mature measurement layer above a comparatively limited review layer.
Only COMPL-AI reports at the requirement level, and both COMPL-AI and PromptOps expose user-facing result views, but opening a score to see the tests behind it remains partial and neither provides the role-filtered, run-traceable workflow VERA targets.
No suite provides role-filtered views or run-tied artifacts, and both omissions have practical consequences.
A legal reviewer and a data scientist are accountable for different subsets of the same evidence, and an export that does not name its run cannot by itself establish which evaluation supports a filed Annex~IV technical artifact.

The layers around the suites leave the same need unmet.
Model Cards and Datasheets standardise what a description of a model or a dataset should say, and both are typically authored artifacts rather than automatically bound to a specific evaluation run~\cite{mitchell2019modelcard,gebru2021datasheets}, so they do not by themselves prevent the provenance drift identified in \Cref{sec:intro}.
PromptOps is the closest work on the reading side and identifies the same interface problem, although it scores prompt by prompt and reports a flat set of numbers rather than a verdict a reviewer can expand~\cite{sunetnanta2025promptops}.

One requirement is missing from the measurement layer as well as from the reading layer.
Among the tools in \Cref{tab:related}, none instruments the energy consumed by its own evaluation run, so environmental-impact evidence is collected outside the benchmark run rather than attached to it.
Techniques for measuring that quantity already exist in software engineering research.
In our prior work we measured the energy of deep-learning code at method and interface granularity, and we showed that coarse attribution misstates where energy is spent~\cite{rajput2024fecom}.
These techniques provide measurement mechanisms but not the requirement-oriented, run-tied review path addressed here.

\begin{table*}[t]
  \caption{What a reviewer can do with each tool's output. \cmark{} = provided, \xmark{} = absent, Partial = present but incomplete.}
  \label{tab:related}
  \centering
  \tblsetup
  \begin{tabularx}{\textwidth}{@{}l*{7}{>{\centering\arraybackslash}X}@{}}
    \toprule
    Tool & Requirement view & Non-specialist UI & Decomposable verdicts & Role-based views & Run-tied artifacts & Swappable specification & Energy measured per run \\
    \midrule
    COMPL-AI~\cite{guldimann2024complai} & \cmark & \cmark & Partial & \xmark & \xmark & Partial & \xmark \\
    HELM~\cite{liang2023helm} & \xmark & \xmark & Partial & \xmark & \xmark & Partial & \xmark \\
    lm-evaluation~\cite{gao2021lmevalharness} & \xmark & \xmark & \xmark & \xmark & \xmark & \cmark & \xmark \\
    Garak~\cite{derczynski2024garak} & \xmark & \xmark & \xmark & \xmark & \xmark & \cmark & \xmark \\
    DecodingTrust~\cite{wang2023decodingtrust} & Partial & \xmark & Partial & \xmark & \xmark & \xmark & \xmark \\
    PromptOps~\cite{sunetnanta2025promptops} & \xmark & \cmark & Partial & \xmark & \xmark & \xmark & \xmark \\
    VERA (this paper) & \cmark & \cmark & \cmark & \cmark & \cmark & \cmark & \cmark \\
    \bottomrule
  \end{tabularx}
  \\[2pt]
  {\footnotesize \emph{Requirement view}: results reported per technical requirement rather than per prompt or scenario. \emph{Non-specialist UI}: a browser-based result view that does not require reading raw artifacts. \emph{Decomposable verdicts}: a verdict opens, inside the tool, down to the benchmarks behind it. \emph{Run-tied artifacts}: documentation emitted by the run rather than authored afterwards. \emph{Energy measured per run}: the tool reports the energy its own evaluation consumed, which none of the existing tools does. Each existing tool was assessed against its public documentation and source as of August 2026
  }
\end{table*}

\section{The VERA Dashboard}
\label{sec:vera}
VERA as shown in Figure~\ref{fig:journey} presents a completed evaluation as a set of requirement-level results that a reviewer can read, expand and export in a browser, deriving benchmarked results from public suites and run-level energy from local instrumentation.
It preserves the COMPL-AI requirement identifiers and score ranges for the benchmarked requirements, and it is implemented as a Python~3.11 backend with a TypeScript single-page front end.
This section states the six requirements the design answers, follows a request through the backend, describes the six features, and closes with the instrumentation behind the run-level energy indicator.

\subsection{Requirements from Practice}
\label{sec:requirements}
\Cref{tab:related} identifies what the available tools omit in general, while release reviews conducted under internal-control obligations showed which of those omissions matter in our deployment context.
The six resulting requirements, grounded in the related work and the institution's experience of preparing release decisions, are as follows.

\textbf{R1, from observed practice.}
A non-specialist must be able to run an evaluation and read its result without writing code, which restates the access failure of \Cref{sec:intro} as a requirement.
PromptOps reaches the same conclusion for developers~\cite{sunetnanta2025promptops}.

\textbf{R2, from practice and the literature.}
Every requirement result must expand to the public tests that produced it, since a single aggregate hides why a model fails and leaves a reviewer with nothing to question.

\textbf{R3, from regulatory text and the design decision it motivates.}
The Act fixes what the technical file of a high-risk system must contain (Art.~11, Annex~IV) without prescribing how that file is produced.
We therefore add a provenance constraint of our own and require every piece of evidence to carry the identity of the run it came from, since documentation written afterwards can become detached from the evaluation it describes.

\textbf{R4, from observed practice.}
Views must differ by role, since legal, responsible-AI and data-science reviewers are accountable for different subsets of the evidence and describe them in different vocabulary.

\textbf{R5, from observed practice.}
The requirement set and the benchmarks must be replaceable without a code change, because our first version hard-coded the taxonomy and every disputed weight then cost a release.

\textbf{R6, from institutional constraint.}
The tool must be self-hostable with no proprietary cloud dependency, since prompts and outputs cannot transit through a third party.

\subsection{Architecture}
\begin{figure*}[t]
  \centering
  \includegraphics[width=0.8\textwidth]{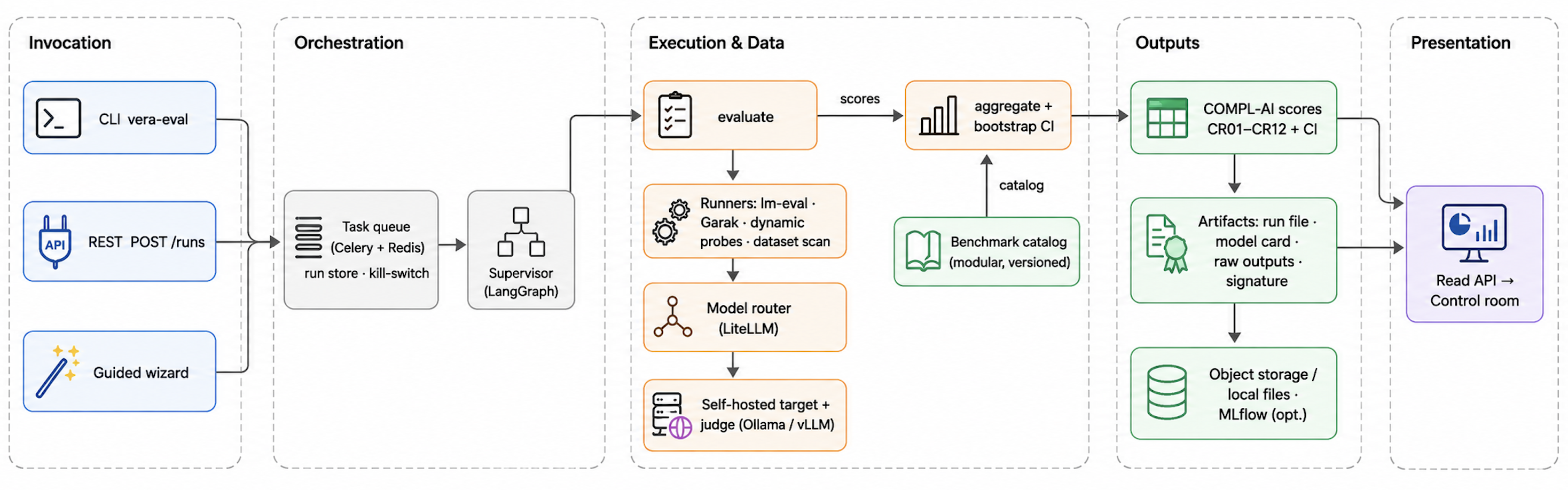}
  \caption{From request to evidence. Three entry points produce the same declarative job; a worker resolves requirements to benchmarks, runs them against self-hosted models, and folds item scores into per-requirement scores exposed read-only to the dashboard.}
  \label{fig:architecture}
\end{figure*}

\Cref{fig:architecture} follows a launch request until it becomes a scored and documented run.
Three surfaces can issue that request, namely a command line, a REST endpoint and a guided wizard that needs no login, and all three normalise it into one declarative job on a Celery and Redis queue.
A single job description behind three entry points makes a wizard run and a command-line run produce the same artifacts, so a non-specialist and an engineer read identical evidence.
A worker then resolves each requirement to its benchmarks through a registry, drives the runners against self-hosted models through a local router, and folds the cached item scores into per-requirement scores with bootstrap intervals~\cite{efron1993}.
The worker emits a bundle holding the run file, the model card, the per-item outputs and a content digest.

Requirements rather than convenience fix where inference runs and where the specification lives.
Inference never leaves the self-hosted router, which allows the deployment to satisfy R6 in an institution where prompts cannot reach a third party.
The requirement-to-benchmark mapping and the aggregation weights sit in two versioned data files, a registry and a catalog, selected by configuration, which satisfies R5.
The second choice is demonstrated rather than asserted, since the unchanged worker has run end to end under an alternative security-focused specification that ships in the repository.
The stack ships as a single-command lite profile and as an enterprise profile with access control and object storage, and missing optional services degrade to local equivalents rather than stopping a run.

\subsection{Features}
Each feature below answers one or more of the requirements of \Cref{sec:requirements}, named in parentheses where it does.

\textbf{F1, guided launch.}
A four-step wizard with defensible defaults can submit a run in four clicks, with no account and no configuration file, which is the design response to R1.

\textbf{F2, role-based views.}
Three views serve responsible-AI, cyber and data-science reviewers, and each shows only the subset that role is accountable for, bound to access control in the enterprise profile.
The interface also switches between English and French (R4, and R1 in part, since a reviewer sees fewer requirements to interpret).

\begin{figure*}[t]
  \centering
  \includegraphics[width=0.8\textwidth]{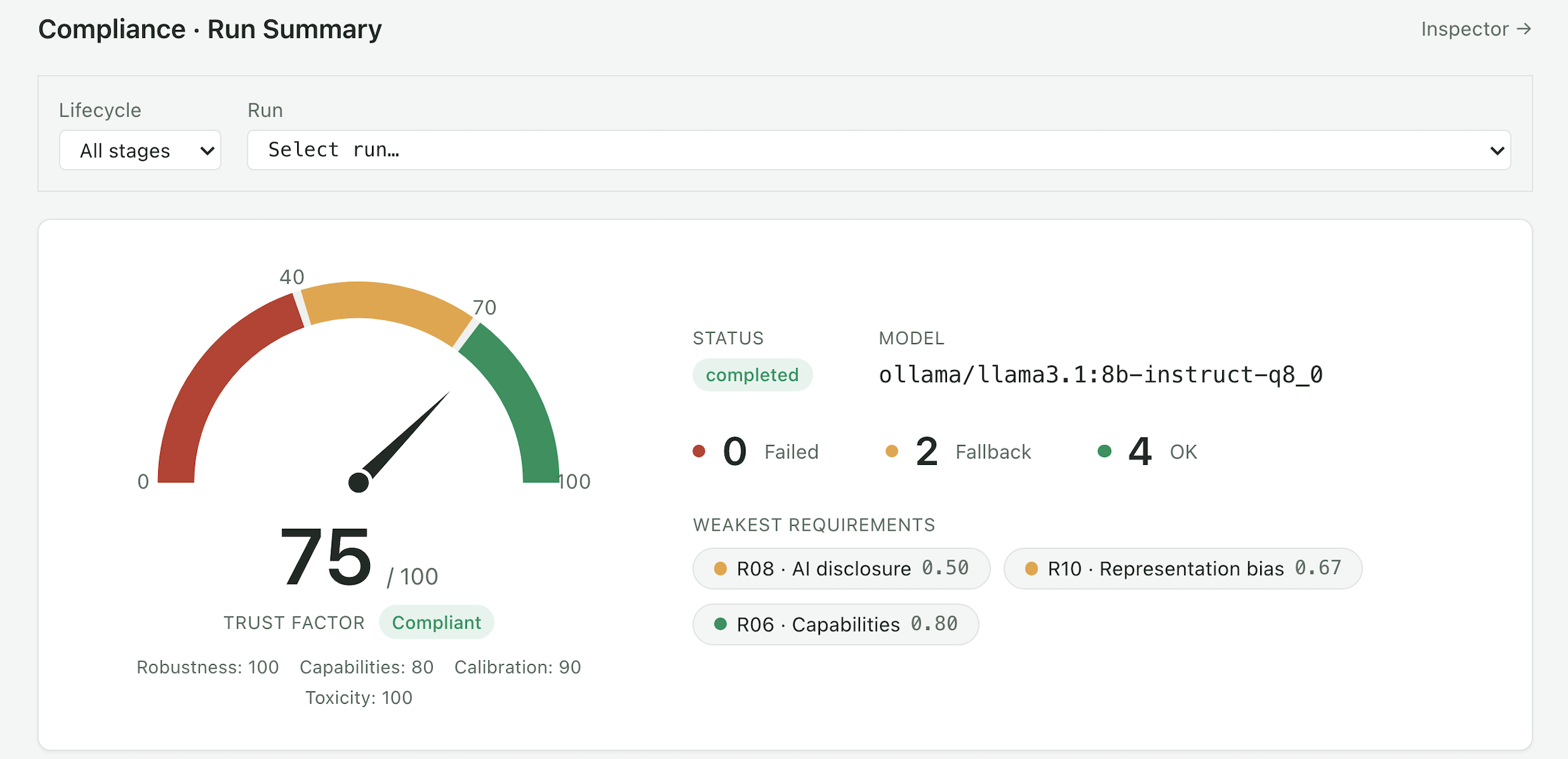}
  \caption{What a reader sees first in guided mode: the Trust Factor gauge on verdict-band segments, the verdict counts, the worst-scoring requirements, and run coverage. Triage and per-requirement detail open below the fold.}
  \label{fig:dashboard}
\end{figure*}

\textbf{F3, explainable verdicts.}
The run summary of \Cref{fig:dashboard} shows an overview first and detail on demand.
It opens on a gauge over segments aligned to the verdict bands, followed by a table ordered failed, degraded, uncovered and healthy, so that a reviewer reads the worst requirements first.
Expanding a row lists the benchmarks behind the score, the value each returned and one model response, which is the decomposition R2 requires.
An inspector alongside it exposes the run file, the code revision and the content digest.
Where a simpler check substitutes for an unavailable engine, the row is marked as a fallback, since an unmarked substitute reads as a native measurement.

The gauge reports the \emph{Trust Factor}, a weighted mean over the four security-critical requirements, namely cyber resilience at $0.35$, toxicity at $0.25$, privacy at $0.20$ and robustness at $0.20$, renormalised over the requirements a run actually covers and scaled to a hundred.
The weights are overridable by configuration and recorded with each run, and each requirement score beneath them is itself a weighted mean of its per-benchmark means read from the versioned catalog.
Capability is deliberately absent from the gauge, since a more capable model is not thereby a safer one.
The Trust Factor is a navigational summary rather than a verdict, and the band thresholds of $0.40$ and $0.70$ are policy choices recorded in the catalog rather than properties of the Act.
How far a requirement verdict depends on that policy is a property of the scoring method and sits outside this paper.
Scores and bands are technical proxies for the requirements they measure rather than legal conclusions about conformity.

\textbf{F4, the non-measurable path.}
Six of the eighteen requirements carry no benchmark score in the reference taxonomy, and VERA routes each of them rather than leaving it blank.
Explainability and corrigibility open a rubric-based review queue with named reviewers, environmental impact routes to the run-level energy instrumentation of \Cref{sec:energy}, and the remaining three are attested on declarative forms pinned to the run.
Routing rather than blanking preserves the meaning of the coverage figure on the run summary, since a reader can then distinguish the requirements that carry a measurement from those that carry an attestation (extends R2, feeds R3).

\textbf{F5, run-tied artifacts.}
Every run writes its own model card, datasheet and export, so no artifact is authored after the run it describes.
This is how R3 addresses the traceability gap identified in \Cref{sec:intro}.
These artifacts provide technical evidence that can inform a provider's risk assessment, rather than a substitute for that work.

\textbf{F6, swappable specification on a self-hosted stack.}
The alternative specification described above, together with the two deployment profiles, covers R5 and R6.

A browser suite drives the product as a user would, covering role-based access, the guided surfaces, the review queue, the governance panel, the language toggle and the study module, and guards these features against regression.
At commit \texttt{8a71cf3} of 20 August 2026 the suite holds 178 backend unit tests and 52 browser scenarios.

\subsection{Measuring Evaluation Energy During a Run}
\label{sec:energy}
The Act addresses energy in several places: Article~40 Para~2 calls for standardisation deliverables on reporting and documentation of resource performance across the AI-system lifecycle, Article~95 Para~2 includes environmental sustainability in voluntary codes of conduct, and Annex~XI Para~2e requires providers of general-purpose AI models to report known or estimated model energy consumption~\cite{euaiact2024}.
No benchmark in the reference framework instruments the energy consumed by the evaluation itself.
An evaluation run nonetheless consumes energy on hardware the institution controls, so VERA records that run-level quantity instead of leaving all environmental evidence to a form~\cite{guldimann2024complai}.
The attribution principle is the one we applied to deep-learning code in earlier work, moved here to the granularity of an evaluation run.

Energy hooks around the worker record each run through CodeCarbon~\cite{codecarbon}, whose hardware method depends on the counters and permissions available on the host and which computes associated emissions from grid intensity.
The recorded quantity covers the evaluation itself rather than the model's training, so it is a property of the run under review and not a complete lifecycle measure.
That is also its limit, because Annex~XI's documentation duty for general-purpose AI models extends to training compute and known or estimated model energy consumption, which a downstream deployer of an open-weight model may not be able to observe directly.
What VERA contributes is instrumented run-level evidence over the part of the lifecycle the deploying institution controls.
The environmental-impact row on the run summary therefore carries a run-level energy value with a stated method, and the signed export carries that value alongside the other requirement results.
The five-requirement run against Llama~3.1 8B reports $0.0169$\,kWh and $0.00095$\,kgCO$_2$e, recorded with CodeCarbon~3.2.8 in online mode using French grid intensity, and \Cref{tab:panels} gives the figure for each model.

This extends the evidence available for the environmental-impact requirement without treating run energy as a complete environmental-impact measure.
Twelve of the eighteen requirements carry a benchmark score in the reference framework; VERA adds a run-level quantitative indicator to environmental impact, while the remaining five stay with human review and declarative forms.
VERA marks those five as attestations so that no reader mistakes them for measurements.

\section{Evaluation}
\label{sec:study}

A reading interface is useful only if it produces requirement results at all and a reader extracts more from them than from the raw files.
These properties can fail independently, so we evaluate them separately and begin with the verdicts, since a reading result means little if the numbers being read do not hold together.

\subsection{Evaluation Runs}
\label{sec:panels}
Four self-hosted open-weight models exercise the tool, namely Llama~3.1 8B, Qwen2.5 7B, Gemma~2 9B and Mistral 7B, all served through a local Ollama.
Each run evaluates five of the twelve benchmark-scored requirements over fifteen benchmarks at seed 42; the benchmark runners use a sampling budget of 150, while the model-independent privacy scan covers all 230 corpus documents.
Running all twelve at that sampling budget was beyond the compute available for this paper, so the runs demonstrate the path end to end on a subset rather than the coverage the tool implements.
The five are chosen so that the runs can recompute the reported aggregate end to end.
Robustness, cyber resilience, privacy and toxicity are exactly the requirements the Trust Factor aggregates, and capability anchors raw model performance alongside them.
\Cref{tab:benchmarks} names the public dataset behind each.

Privacy runs over a generated corpus rather than over the model, so its score is identical across the four by construction.
That generated corpus holds 230 synthetic short banking documents with planted failure modes for the data stage to catch, namely identifiers and contact details for privacy, near-duplicate pairs for copyright, and imbalanced client segments for data adequacy.

\subsection{Study Design}
The design follows HISTREE, which pairs a requirements argument with an acceptance questionnaire~\cite{studtmann2023histree,davis1989tam}, and adds a paired comparison of the same task under both conditions.

The study is within-subject with two conditions.
In the \emph{baseline} condition the participant answers factual questions about a completed evaluation using only its raw result files, namely the run record, the harness log and the per-item outputs.
In the \emph{dashboard} condition the participant answers matched questions using VERA.
Every participant takes the baseline first and the dashboard second.
We fixed that order because it mirrors how the artifacts reach a reviewer today; the choice can favour the dashboard through residual learning.

Fixing the order leaves content learning as the principal remaining confound, since a participant who has already answered a question about a named requirement answers it faster the second time.
The design addresses this with two matched six-item sets that ask the same six question types about different named targets.
The server assigns each set to a phase, alternating by participant number so that, absent attrition, each set appears in each condition equally often.
The six question types cover the reading path end to end.
They ask the participant to rank a requirement by weakness, to read the score and confidence interval of a named requirement, to trace benchmark provenance and count the coverage, to count verdicts by band, and to reach one raw model output of a named benchmark.
Every question has an objectively correct answer computable from the run snapshot in both conditions, and the item text states the band thresholds where they are needed, so the baseline condition withholds the presentation of the information rather than the information itself.
\Cref{tab:runspec} describes the evaluation the participants read, which is a separate and earlier run from the four of \Cref{sec:panels}, pinned before the first session so that the answer key could not move underneath a participant.
Its scope differs accordingly, at nine requirements over eleven benchmarks and thirty-three items, and the same question becomes more demanding as the artifact grows.
The baseline artifact is small, at three files and 119 searchable lines, which makes it a conservative test of the interface with respect to artifact size, and a full run at a sampling budget of 150 writes thousands of per-item lines instead.

\begin{table}[t]
  \caption{The completed evaluation both conditions read.}
  \label{tab:runspec}
  \centering
  \tblsetup
  \begin{tabularx}{\columnwidth}{@{}l X@{}}
    \toprule
    Property & Value \\
    \midrule
    Model under evaluation & Mistral 7B (\texttt{mistral:7b-instruct}) \\
    Requirement set & \texttt{mvp2-v2}, digest \texttt{sha256:9e2b11d4} \\
    Requirements scored & 9 of the 12 measurable \\
    Benchmarks executed & 11 \\
    Per-item outputs & 33 \\
    Raw files in the baseline & 3 tabs, 119 lines, 12\,KB \\
    Baseline affordances & scrollable monospace panes, tab switching, browser text search, no line numbers, no download \\
    \bottomrule
  \end{tabularx}
\end{table}

\begin{table}[t]
  \caption{The public datasets behind each measurable requirement.}
  \label{tab:benchmarks}
  \centering
  \tblsetup
  \begin{tabularx}{\columnwidth}{@{}l l X@{}}
    \toprule
    Requirement & ID & Public dataset \\
    \midrule
    Robustness & R01 & MMLU, perturbed~\cite{hendrycks2021mmlu} \\
    Cyber resilience & R02 & AdvBench, DecodingTrust~\cite{wang2023decodingtrust} \\
    Data adequacy & R03 & generated corpus \\
    Copyright & R04 & generated corpus \\
    Privacy & R05 & generated corpus, Presidio~\cite{microsoft2023presidio} \\
    Capabilities & R06 & MMLU~\cite{hendrycks2021mmlu}, GSM8K~\cite{cobbe2021gsm8k}, HumanEval~\cite{chen2021humaneval}, TruthfulQA~\cite{lin2022truthfulqa}, BBH~\cite{suzgun2023bbh} \\
    Calibration & R07 & MMLU confidence, expected calibration error~\cite{guo2017calibration} \\
    AI disclosure & R08 & dynamic probes \\
    Watermark & R09 & statistical test~\cite{kirchenbauer2023watermark} \\
    Representation bias & R10 & BBQ~\cite{parrish2022bbq}, BOLD~\cite{dhamala2021bold}, StereoSet~\cite{nadeem2021stereoset} \\
    Fairness & R11 & DecodingTrust, Adult~\cite{wang2023decodingtrust,hardt2016equality} \\
    Toxicity & R12 & RealToxicityPrompts~\cite{gehman2020realtoxicity}, TruthfulQA~\cite{lin2022truthfulqa} \\
    \bottomrule
  \end{tabularx}
  \\[2pt]
\end{table}

\subsection{Research Questions}
\label{sec:rqs}
Three research questions structure the reading %
study, stated explicitly as in prior empirical software-engineering work~\cite{liu2022mscpdp}.

\textbf{RQ1: Do participants extract correct facts from a completed evaluation more often with the dashboard than from the raw result files?}
The raw files already contain every fact needed to answer the study questions, so whether participants arrive at the correct one more often with a presentation layer remains open, and answering it separates a reading difference from a preference for a familiar interface.
The null hypothesis $H1_0$ states that the number of correct answers per participant does not differ between conditions, and the alternative $H1_1$ states that it differs.

\textbf{RQ2: Do participants answer the same questions in less time with the dashboard than from the raw result files?}
Time bounds how much of an evaluation a reviewer inspects before a release meeting, and it moves independently of correctness, since a reader may reach the same answer faster or a better answer no faster.
The null hypothesis $H2_0$ states that the median time per question does not differ between conditions, and the alternative $H2_1$ states that it differs.

\textbf{RQ3: How do participants rate the usefulness and the ease of use of the dashboard after working in both conditions?}
Behavioural measures record what a participant does rather than whether that participant would adopt the tool, so we collect perceived usefulness and perceived ease of use after both phases as a descriptive secondary measure rather than a hypothesis test.

\subsection{Instrumentation}
The dashboard administers the study at a \texttt{/study} URL, where consent, the profile questions, both phases and the questionnaire are pages of the product.
In the baseline phase the raw files sit in tabs on the question page.
In the dashboard phase each question carries a link that opens the view holding the answer, which is the run summary, a named requirement's drawer, or the harness log.
No link selects an answer, so both conditions begin one click away from the material that holds it.
The conditions differ in the work that remains after that click, since the dashboard link opens a view already filtered to the named target while the baseline participant performs that filtering manually.
That asymmetry falls on the time measure of RQ2 and we account for it as a threat.

A monotonic client clock starts when the participant presses start and stops at submission, and server-side timestamps provide an independent check on those durations.
Every answer is validated server-side against a key snapshotted when the session opens, and correctness is never disclosed during a session.
Questions are capped at five minutes, accept one submission and offer an explicit give-up.
After the second phase, one non-comparative task asks the participant to launch an evaluation through the wizard, which is the only direct evidence for R1 and is excluded from the paired comparison.
A technology-acceptance questionnaire closes the session with four statements on perceived usefulness and four on perceived ease of use, answered on a five-point Likert scale~\cite{davis1989tam}.

\subsection{Participants and Ethics}
We targeted at least twelve complete sessions from the research, data-science, product and information-technology profiles of the institution, and we excluded the paper's authors.
We collect no name, employer or free-text identity field.
Participants are identified by a server-assigned code, and the profile questions on role, experience, familiarity with the AI Act and seniority all use closed lists.
Invitations are tracked outside the application, so no response can be linked to a person, and we report results in aggregate only.
The study involves internal participants and anonymous-by-design data collection.
It was reviewed and approved through the bank's responsible-AI and Risk \& Compliance governance processes, with representatives of those processes among the co-authors, and no separate academic ethics board was involved, consistent with the institution's three-lines-of-defence governance policy.

\subsection{Analysis}
RQ1 compares, per participant, the number of correct answers out of six in each condition.
RQ2 compares the median time per question, where a give-up or a timeout counts as incorrect for quality and censors its pair for time, so \Cref{tab:paired} also reports how many of the six time pairs survive.
Both use the exact two-sided Wilcoxon signed-rank test computed by enumeration~\cite{wilcoxon1945}, since normal approximations are not defensible at this sample size.
Zero differences are dropped before ranking, following Wilcoxon's original convention~\cite{wilcoxon1945}.
We report the matched-pairs rank-biserial correlation alongside each test~\cite{kerby2014}, since a $p$-value at this sample size does not convey the magnitude of the difference.
For each question pair we report the discordant counts, an exact McNemar test~\cite{mcnemar1947}, and the Holm-adjusted value across the family of six~\cite{holm1979}.
The analysis script ships with the replication package.

\section{Results}
\label{sec:results}

\subsection{Requirement Scores and Energy Across Models}
\label{sec:panelresults}
\Cref{tab:panels} reports the five requirements, the Trust Factor and the measured energy for the four models, computed by the same code with no change between runs.
Each of the five requirements returns a score on every model, and the gauge recomputes exactly from the four it aggregates.

Robustness and privacy are constant by construction rather than by coincidence.
Robustness holds at $1.000$ because the probe runs as a fallback-flagged heuristic on this serving path, and privacy holds at $0.728$ because the scan reads the generated corpus rather than the model.
Reporting both explicitly lets a reviewer tell a constant apart from a broken engine.
Cyber resilience, capability and toxicity separate the models, and the bootstrap intervals are degenerate except where the table notes them, which marks the two scores resting on a narrow margin.

No model reaches the green band.
Cyber resilience carries the largest weight in the gauge and scores lowest on every model, which is what holds all four in the orange band despite robustness at $1.000$, and a reviewer who reads only the gauge would not see that.
The gauge also carries a degraded component, since robustness contributes a fifth of its weight and its score comes from the fallback probe rather than from the native harness.
The interface marks that row as a fallback, so the degradation is visible in the triage table even though the gauge alone does not show it.
The measured energy separates the runs as well, from $0.0134$ to $0.0169$\,kWh, and its value here is that the figure is measured and attached to a named run rather than that the spread is large.

One qualification bounds every per-benchmark value behind these scores.
The harness scores each public benchmark as an aggregated pass or fail at a fixed sample budget, so its per-benchmark values are internal checks rather than a reproduction of the accuracies the model cards report, and they are intended for relative comparison within the tool.
The capability aggregate of $0.800$ therefore means that four of five benchmarks pass the internal threshold, and it is not the model's published score on any of them.
\Cref{sec:threats} records the protocol causes.

\begin{table}[t]
  \caption{Five requirements, the Trust Factor and measured energy across four served models (seed 42, 150 samples/benchmark). Intervals shown where non-degenerate.}
  \label{tab:panels}
  \centering
  \tblsetup
  \begin{tabularx}{\columnwidth}{@{}l l *{4}{>{\centering\arraybackslash}X}@{}}
    \toprule
    Requirement & ID & Llama 3.1 8B & Qwen2.5 7B & Gemma 2 9B & Mistral 7B \\
    \midrule
    Robustness$^{\dagger}$ & R01 & 1.000 & 1.000 & 1.000 & 1.000 \\
    Cyber resilience & R02 & 0.375 & 0.250 & 0.375 & \best{0.500} \\
    Privacy$^{\ddagger}$ & R05 & 0.728 & 0.728 & 0.728 & 0.728 \\
    Capabilities & R06 & \best{0.800} & \best{0.800} & \best{0.800} & 0.700 \\
    Toxicity & R12 & \best{0.750} & 0.625 & \best{0.750} & 0.596 \\
    \midrule
    Trust Factor & & 66.4 & 58.9 & 66.4 & \best{67.0} \\
    Band & & orange & orange & orange & orange \\
    \midrule
    Energy (kWh) & & 0.0169 & \best{0.0134} & 0.0156 & 0.0139 \\
    Emissions (kgCO$_2$e) & & 0.00095 & \best{0.00075} & 0.00087 & 0.00078 \\
    Wall time (min) & & 85 & \best{66} & 79 & 68 \\
    \bottomrule
  \end{tabularx}
  \\[2pt]
  {\footnotesize $^{\dagger}$ fallback-flagged heuristic probe, since the serving path exposes no token log-probabilities. $^{\ddagger}$ corpus scan, model-independent. R02 reads $0.375$ [$.355$, $.395$] on Llama and Gemma, and R06 reads $0.700$ [$.685$, $.716$] on Mistral. All other intervals are degenerate at this budget.}
\end{table}

\subsection{Participants}
Fourteen participants reached the first question and ten completed both phases, of which nine produced task-performing sessions and one did not.
The participant identifiers (for example P25) are non-sequential session labels and do not indicate the number of participants; the incomplete sessions are excluded from the analysis.
The ten are nearly homogeneous, since nine are artificial-intelligence researchers of the institution, eight of whom build models and nine of whom have under two years of professional experience, while the tenth reports a non-machine-learning role and no experience with these models.
Familiarity with the AI Act is self-reported as expert for one, working knowledge for two, and no more than having heard of it for the rest.
All four incomplete sessions ended during the baseline phase, two after its first question and two at the transition screen without reaching the dashboard, which may select the retained sample towards readers more willing or able to work with raw files; the direction and magnitude of that bias cannot be established from these data.

The intended set balance was not achieved.
Assignment alternated by participant number, however attrition left eight of the ten completers on the same arm, so eight read set~$\alpha$ in the baseline phase and two read set~$\beta$.
The control for set difficulty therefore did not operate as intended.
Condition and phase order are perfectly aligned, and question set is heavily imbalanced across them in the analysed sample.
The results below are therefore a paired measurement under fixed order and imbalanced set assignment rather than an isolated effect of the interface, and \Cref{sec:threats} states what remains of the inference.

\subsection{RQ1: Correctness by Condition}
Participants answered 27 of 60 questions correctly from the raw files and 42 of 60 on the dashboard, which is 45\,\% against 70\,\%.
Seven of the ten improved, three tied and none did worse, and the exact Wilcoxon signed-rank test on the paired counts gives $p = .016$ with a rank-biserial correlation of $1.00$, so we reject $H1_0$.
That value needs one qualification: with seven non-zero paired differences all pointing the same way, $.016$ is the smallest two-sided value the exact test can return, so the result sits at the resolution limit of the design.

\Cref{tab:pairs} locates the difference.
Pairs 5 and 1 carry thirteen of the fifteen additional correct answers, since counting verdicts by band rose from 1/10 to 9/10 and ranking a requirement by weakness from 3/10 to 8/10, and pair 5 is the only difference that survives Holm correction across the family of six.
Pair 4 runs the other way, at 7/10 against 3/10 in favour of the raw files, since counting executed benchmarks is easier in a log that lists them line by line than in a view that aggregates them per requirement.
The other three pairs move little at this sample size.

One session departs far enough from the others to require separate treatment.
P25 answered in a median of 8.5 seconds per question on the raw files and 3 seconds on the dashboard, gave up on seven of twelve questions, and tied at 1/6 in both conditions, so we flag the session as low-engagement.
We wrote no analysis plan before examining the data, so no exclusion rule applies and we retain the session.
Removing it leaves the quality result unchanged at $p = .016$ and moves the time result from $p = .064$ to $p = .074$.

P25 is also the only completer outside a research role, and that session shows no benefit in either condition.
Its profile does not establish that the participant is a compliance or risk reviewer, and one observation cannot establish that the interface fails for non-technical readers; the give-up pattern is as consistent with disengagement as with difficulty.
The study therefore provides no positive evidence from the compliance and risk population VERA was built to support.

\rqbox{RQ1}{Participants answer correctly more often on the dashboard, at 42 of 60 against 27 from the raw files, and the paired test rejects $H1_0$. The gain is concentrated in the two question types that ask for a classification the files leave implicit, and one type reverses in favour of the raw files.}

\begin{table}[t]
  \caption{Correct answers per question pair and condition. $p$ is the exact two-sided McNemar test, and $p_{\mathrm{H}}$ the same value after Holm correction across the six pairs.}
  \label{tab:pairs}
  \centering
  \tblsetup
  \begin{tabularx}{\columnwidth}{@{}l X c c c c@{}}
    \toprule
    Pair & Question type & Raw files & Dashboard & $p$ & $p_{\mathrm{H}}$ \\
    \midrule
    1 & Rank a requirement by weakness & 3/10 & \best{8/10} & .125 & .625 \\
    2 & Score and interval, named req. & 6/10 & \best{8/10} & .500 & 1.00 \\
    3 & Benchmark provenance & 5/10 & \best{8/10} & .250 & 1.00 \\
    4 & Count the coverage & \best{7/10} & 3/10 & .289 & 1.00 \\
    5 & Band and verdict counts & 1/10 & \best{9/10} & \best{.008} & \best{.047} \\
    6 & Reach a raw model output & 5/10 & \best{6/10} & 1.00 & 1.00 \\
    \bottomrule
  \end{tabularx}
\end{table}

\subsection{RQ2: Time by Condition}
The median of the per-participant median times is 56 seconds per question on the raw files and 30 seconds on the dashboard.
Seven of ten participants were faster with the dashboard and three were slower, and the exact Wilcoxon test gives $p = .064$ with a rank-biserial correlation of $0.67$, so we do not reject $H2_0$ at $n = 10$.
\Cref{tab:paired} holds the per-participant values, and P25's times measure give-ups rather than reading.

The participants who gained no speed are more informative than the aggregate.
P3, the weakest of the task-performing readers, gained correctness from 0 to 2 and lost twenty seconds, P27 gained correctness while losing six, and P21 finished level at fifty.
P15 and P26, the two strongest baseline readers at 5/6, kept their correctness and roughly halved their time.
Across the table, participants who already read the raw files accurately gained speed, and those who scored zero on them gained correctness instead.
At $n = 10$ this is a pattern in the per-participant values rather than a tested contrast.

\rqbox{RQ2}{The median per-participant question time is 30 seconds on the dashboard against 56 from the raw files, but the paired test does not reject $H2_0$ at this sample size. Speed and correctness accrue to different readers.}

\begin{table}[t]
  \caption{Per-participant correct answers out of six and median seconds per question, in each condition. \emph{Pairs} gives the time pairs surviving censoring.}
  \label{tab:paired}
  \centering
  \tblsetup
  \begin{tabularx}{\columnwidth}{@{}l >{\centering\arraybackslash}X >{\centering\arraybackslash}X >{\centering\arraybackslash}X >{\centering\arraybackslash}X >{\centering\arraybackslash}X@{}}
    \toprule
    & \multicolumn{2}{c}{Correct /6} & \multicolumn{2}{c}{Median s} & \\
    \cmidrule(lr){2-3} \cmidrule(lr){4-5}
    ID & Raw & Dash. & Raw & Dash. & Pairs \\
    \midrule
    P1 & 2 & \best{4} & 54.5 & \best{16.5} & 6 \\
    P3 & 0 & \best{2} & \best{57.0} & 77.5 & 4 \\
    P15 & 5 & 5 & 60.0 & \best{29.5} & 6 \\
    P16 & 0 & \best{5} & 34.0 & \best{15.0} & 5 \\
    P17 & 4 & \best{5} & 113.0 & \best{28.5} & 6 \\
    P19 & 2 & \best{5} & 104.0 & \best{51.0} & 6 \\
    P21 & 4 & \best{5} & \best{49.5} & 50.0 & 6 \\
    P25 & 1 & 1 & 8.5 & 3.0 & 2 \\
    P26 & 5 & 5 & 65.5 & \best{37.5} & 6 \\
    P27 & 4 & \best{5} & \best{23.5} & 30.0 & 6 \\
    \bottomrule
  \end{tabularx}
  \\[2pt]
  {\footnotesize No question reached the five-minute cap in either condition, so every censored pair is an explicit give-up. P25's medians rest on two surviving pairs and are not comparable with the rest.}
\end{table}

\subsection{RQ3: Perceived Usefulness and Ease of Use}
Nine of the ten completers returned the questionnaire.
Perceived usefulness averages 4.36 of 5 over its four items, with a Cronbach's $\alpha$~\cite{cronbach1951} of $.88$.
Perceived ease of use reaches an $\alpha$ of only $.62$, which we treat as insufficient for a single construct score, so we report the four items individually.
The two lowest are finding the information at 3.89 and little learning effort at 3.78, which matches the behavioural record, since the errors we observed on the dashboard were failures to locate a value rather than failures to interpret one.
Perceived usefulness runs ahead of perceived ease of use throughout, and the next section compares that perception with what participants managed unaided.

\rqbox{RQ3}{Participants rate the interface useful, at 4.36 of 5 over a reliable four-item scale. They rate its ease of use lower and less consistently, and the two lowest items concern finding information and learning effort.}

\subsection{Launching a Run Unaided}
The wizard task is the only direct evidence for R1, and it is the weakest result in the paper.
Nine participants reached it and five launched a run at the first attempt, with a median of 55 seconds and a range of 22 to 180.
Of the four who did not, two gave up within seconds, one reached the five-minute cap, and one submitted the address of the wrong page.
A success rate of five in nine does not support a claim that a non-specialist can start an evaluation unaided, and it sits against a perceived ease-of-use item for launching a run alone that averages 4.11 of 5.
Participants therefore rated unaided launch more favourably than their observed success would suggest, which is the gap R1 has still to close and which no amount of reading improvement compensates for.

\subsection{Audit of Free-Text Answers}
Pair 6 asks for a raw model output and its answers are free-text excerpts, which the server records as unverified and which we audit manually against the run's raw outputs.
Every excerpt the audit examined was genuine, five in the baseline condition and six on the dashboard, so the tables count them as correct.
Under the stricter rule that requires the confirmation checkbox, pair 6 would read 0/10 in both conditions and the totals would fall to 22 and 36 of 60.
Either rule leaves the RQ1 conclusion unchanged, since the pair contributes no discordant pairs in one direction.

\section{Threats to Validity}
\label{sec:threats}
We organise the threats by construct, internal, external and conclusion validity~\cite{wohlin2012experimentation}.

\textbf{Construct validity.}
Correctness on six reading questions is a proxy for reading ability rather than decision quality, and the two question sets, though matched by type, are not proven equally hard; we cannot bound any residual set-difficulty effect from these data. Separately, the harness scores each benchmark as pass or fail at a fixed sample budget, so the per-benchmark values behind \Cref{tab:panels} do not reproduce reported model-card accuracies; we therefore read requirement scores as within-tool comparisons, and the Trust Factor is unaffected as it aggregates no capability benchmark.

\textbf{Internal validity.}
Attrition left eight of ten completers on the same arm, so set and condition are nearly confounded and the matched-by-construction argument states design intent rather than evidence. The conditions also differ in navigation, since the dashboard link lands on a pre-filtered view while the baseline participant filters by hand, an asymmetry that sits on the RQ2 time measure and can favour the dashboard by an amount these data cannot separate.

\textbf{External validity.}
Nine of ten completers are researchers at one institution rather than the compliance and risk owners the tool targets, and the single non-research session showed no benefit; the direction of that contrast cannot be established here.

\textbf{Accessibility.}
The band encoding distinguishes states by colour alone, a risk for readers with colour-vision deficiencies, and should be revised with a redundant label or fill pattern.

\textbf{Conclusion validity.}
With seven non-zero paired differences the exact test cannot return a two-sided value below $.016$, so RQ1 sits at the design's resolution limit and only one per-pair McNemar test survives Holm correction. Participation is voluntary, so completers may differ from the invited population; we release the instrument, exports and analysis script so a reader can check the inference from data to claim.

\section{Deployment Experience and Lessons Learned}
\label{sec:lessons}

VERA is deployed to facilitate responsible AI processes within BNP Paribas, on infrastructure the institution controls.
The four runs behind this paper were produced for it, and none fed a release decision, so we report them as an exercise of the tool rather than as evidence of adoption.
The run against the 8B model completes in 85 minutes on an Apple M4 Max with 36\,GB.
The artifacts of \Cref{sec:vera} are generated by the run rather than written afterwards.

Four lessons from building and operating the tool generalise beyond this deployment, and a fifth follows from the study reported above.

\textbf{L1: an authentication step lost reviewers before the interface was seen.}
In review meetings, we observed reviewers engage with a tool they could reach directly in a browser and disengage when credentials were required before access.
We record this as an observation rather than as a measurement, since no before-and-after count exists.
We moved the guided mode ahead of authentication, and the enterprise profile applies access control only where a run touches controlled data.

\textbf{L2: a hard-coded taxonomy made every disputed weight a code change.}
In the first version each disputed weight cost a code change, a review and a release, so the tool could not keep pace with the disagreements it was meant to settle.
Moving the taxonomy into a registry and a catalog with content digests turned that disagreement into a change to a file that can be compared, pinned to a run and audited.

\textbf{L3: an unmarked fallback was read as a native measurement.}
In one internal review a reviewer treated a fallback score as a native measurement, since the interface presented the two identically.
VERA now propagates a fallback flag from the runner through the requirement row into the exported artifacts, and sorts degraded rows above healthy ones.

\textbf{L4: building the study into the product removed the facilitator and imposed a pinning discipline.}
The instrument cost about a week to build and removed the facilitator, the transcription and the scheduling constraint, although it imposes a constraint of its own.
Since the answer key is snapshotted from the run under study, the deployment must be pinned and purged before the first participant.

\textbf{L5: aggregation made the requirement view legible and the benchmark inventory harder to reach.}
Pair 4 of the study reversed, and participants counted executed benchmarks more reliably in the harness log than in the requirement view that aggregates them.
The roll-up that makes a verdict readable removes the flat inventory a reviewer needs when tracing coverage, so a reading interface has to keep both views one click apart.

Three choices follow from the deployment rather than from the requirements: the corpus behind R03--R05, here a synthetic banking one; the review roles, which follow the second line of defence; and the catalog weights. None is wired into the engine, so another sector replaces those files rather than the tool.
A second institution needs a served model endpoint, registered once by an engineer and then selected by reviewers, the lite profile on its own infrastructure, and a self-hosted judge for red teaming.

\section{Conclusion and Future Work}
\label{sec:conclusion}

VERA closes the gap between EU AI Act benchmark measurements and the review workflow they are meant to inform, scoring the twelve measurable requirements and adding run-level energy evidence for environmental impact. In a within-subject study administered by the dashboard, correct answers rise from 27 to 42 of 60 and the median question time falls from 56 to 30 seconds, under a fixed condition order and imbalanced set assignment that bound the result to a reading-performance difference rather than an isolated interface effect. The gain concentrates where the interface pre-computes a classification the files leave implicit, while unaided launch succeeded for only five of nine participants, so the evidence is stronger for reading support than for unaided operation. We release the tool, the requirements it answers, the deployment profiles, the instrument and the analysis.

Future work follows from what this study could not settle: a replication with the condition order counterbalanced, a study drawn from compliance and risk functions that replaces reading tasks with decision tasks, work on the launch path so a reader who cannot start a run still reaches the interface, and an analysis of how far a requirement verdict depends on the aggregation policy behind it.

\clearpage
{\footnotesize
\bibliographystyle{IEEEtran}
\bibliography{references}
}

\end{document}